\documentclass[aps,twocolumn]{revtex4}
\usepackage{latexsym}
\usepackage{epsfig}
\usepackage{graphicx}

\begin{document}
\title{Vibrational modes identify soft spots in a sheared disordered packing}
\author{M. L. Manning}
\affiliation{Princeton Center for Theoretical Science, Princeton, NJ 08544}
\email{lm2@princeton.edu}
\author{A. J. Liu}
\affiliation{Department of Physics, University of Pennsylvania, Philadelphia, PA 19130}
\begin{abstract}
We analyze low-frequency vibrational modes in a two-dimensional, zero-temperature, quasistatically sheared model glass
to identify a population of structural ``soft spots'' where particle rearrangements are initiated.  The population of spots evolves slowly compared
to the interval between particle rearrangements and the soft spots are structurally different from the rest of the system.  Our results suggest 
that disordered solids flow via localized rearrangements that tend to occur at soft spots, which are analogous to dislocations in crystalline solids.
\end{abstract}

\maketitle

Like liquids, solids can flow under applied shear stresses.  Crystalline solids flow via rare rearrangements controlled by a population of lattice defects, namely dislocations~\cite{Taylor}.  In disordered solids, rearrangements tend to be localized~\cite{Gopal,Falk_L1,Schall} but there is no obvious way to identify defects that might control them~\cite{GILMAN,CHAUDHARI}.  Can these rare localized rearrangements occur anywhere, as in a liquid, or do glasses possess a population of ``soft spots," analogous to dislocations in crystalline solids, which are structurally distinct and susceptible to rearrangement?  Although useful continuum models assume the latter~\cite{Falk_L1, SGR}, such a population of spots has never been identified from structural information. 

In order to search for a population of soft spots, we must start with a solidlike description of the glass.  We begin with harmonic theory, in which the linear response to an applied stress is completely characterized by the normal modes of vibration.  This approximation breaks down before solids begin to flow, so one would not expect the linear response to yield much insight into particle rearrangements.  However, recent evidence suggests that low-frequency vibrational modes, which are generically more prevalent in disordered solids than in crystalline ones~\cite{Silbert,Wyart2,LiuNagel2010}, can be quasilocalized.  Such modes have unusually low energy barriers to rearrangements~\cite{Ning}, and therefore are correlated with rearrangements~\cite{WidmerCooper,Brito1,Tanguy3,Tanguy4}. 

In this paper, we use low-frequency modes to identify a population of soft spots in a model glass.    We find that  rearrangements begin at soft spots, the population of soft spots evolves slowly compared to the time between rearrangements, and that there are structural differences between soft spots and the rest of the system.   We therefore conclude that soft spots are good candidates for elementary defects that control the flow of disordered solids.

We study a 50:50 binary mixture of soft discs of diameter ratio $1.4$ in two dimensions, interacting via a Hertzian potential $V = \epsilon (1 -r/R)^{5/2}$, where $r$ is the distance between the centers of two particles and $R$ is the sum of their radii.  Results presented here are for jammed packings with a packing fraction $\phi=0.95$, which is much higher than the jamming transition at $\phi_c \simeq 0.84$.  We have also identified soft spots at values of $\phi$ closer to the transition~\cite{lisa}.  Lengths and frequencies are in units of the small particle diameter and the interaction energy $\epsilon$. We employ Lees-Edwards boundary conditions to shear the system with a strain step of $10^{-5}$.  After each strain step we relax the structure to its minimum energy to shear the system athermally and quasistatically.  

With increasing applied strain, the shear stress increases with a slope given by the shear modulus, then drops abruptly when there is a rearrangement.  The strain step size is reduced to $2 \times 10^{-7}$ before each rearrangement.  Between rearrangements, the dynamical matrix $M$ is calculated at small strain intervals to obtain its eigenvalues (corresponding to the square of the frequency) and eigenvectors (the vibrational modes)~\cite{ARPACK}. 

In this limit of zero temperature and strain rate, a rearrangement occurs when one vibrational mode (the critical mode) reaches zero frequency.   At that critical strain, $\gamma_c$, the packing becomes unstable and the coordination of particles in the packing changes.  The initial rearrangement can trigger an avalanche of additional particle motions~\cite{Maloney2}, so that the net displacements of the particles may be very different from the critical mode and may involve contributions from a number of modes~\cite{Tanguy4}.   However, in solids at finite temperatures and strain rates, fluctuations can interrupt or extend avalanches.  Therefore, for the remainder of this paper we focus not on the avalanche but on the reproducible {\em initial} particle rearrangement, described by the critical mode~\cite{Maloney2}.    

\begin{figure}\centering \includegraphics[width=0.40\textwidth]{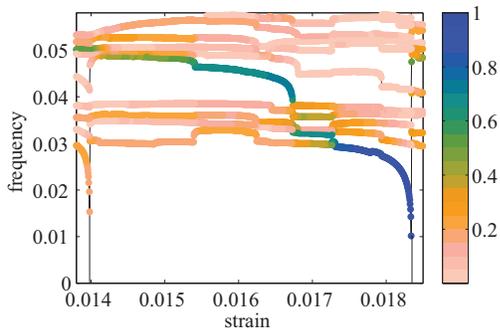}
\caption{ \label{evstrain} (color online) The lowest ten normal mode frequencies as a function of applied strain.  There are two critical strains at which a mode frequency approaches zero and particles rearrange, at $\gamma_c \simeq 0.014$  and $\gamma_c \simeq 0.0183$.  The color of each point indicates the overlap of that mode with the critical mode at $\gamma_c \simeq 0.0183$.  The lowest energy mode does not resemble the critical mode until just before the particle rearrangement.} 
\end{figure}

As the system is strained, the packing becomes less stable and the mode frequencies tend to shift downwards.  At a given strain, one might expect the lowest frequency mode to be the one whose frequency vanishes at the next rearrangement.  Fig.~\ref{evstrain} shows this is not generally true -- the mode most similar to the critical mode lies at the lowest frequency only for a small range of strains immediately preceding the particle rearrangement~\cite{Tanguy4}.  Note that most excitations at low frequencies are weakly-scattered sound waves with a strong plane-wave character.  These excitations coexist in the same frequency range as the quasilocalized excitations~\cite{Ning}; as a result, the normal modes exhibit characteristics of each.  We therefore look at the entire population of low-frequency modes to extract soft spots, as follows.

\begin{figure}
\centering \includegraphics[width=0.47\textwidth]{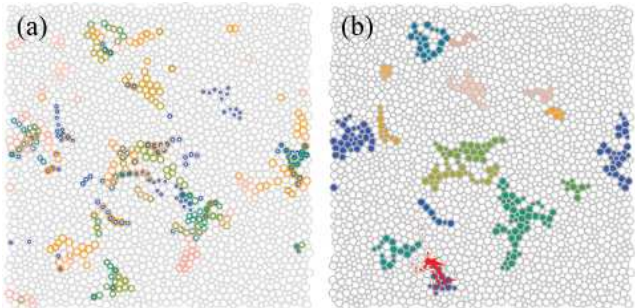}
\caption{ \label{sspots} (color online) Soft spots in a system calculated at $3.2 \times 10^{-3}$ units of strain before a particle rearrangement.  (a) Regions of large displacement in the$N_m = 30$ lowest frequency modes. Bold circles highlight the $N_p = 20$ particles with the largest polarization vectors, and different colors correspond to different modes.  (b) Soft spots generated by clustering the particles highlighted in (a). Red arrows indicate the displacement of each particle during the next rearrangement.}
\end{figure}

For a granular packing of $N=2500$ particles, we first identify the $N_m$ lowest frequency modes in the spectrum of the dynamical matrix and the $N_p$ particles in each of these modes with the largest polarization vectors. The values of $N_m$ and $N_p$ are eventually varied to maximize correlation with particle rearrangements.  Fig.~\ref{sspots}(a) illustrates the locations of the particles identified by the lowest $N_m = 30$ modes and $N_p = 20$ particles for a particular configuration.  Note that the largest polarization vectors are spatially clustered into regions, and that the same regions appear in several different modes.   Each of the $N_p$ particles in each of the $N_m$ modes is then assigned a value of unity, while the remaining particles are assigned a value of zero.  We separate this binary map into localized clusters or ``soft spots" as shown in Fig.~\ref{sspots}(b)~\cite{remark}.  Thus, the population of soft spots at strain $\gamma$ is represented by a binary vector ${\bf S} (\gamma)=\{S_i(\gamma) \in \{ 0,1\} \}$, where $S_i=1$ if particle $i$ is in a soft spot and $S_i=0$ otherwise.  In addition, we construct a binary vector for each soft spot, indexed by $\alpha$: ${\bf s}_{\alpha}= \{s_{\alpha,i}(\gamma) \in \{0,1\} \}$, with $s_{\alpha,i}=1$ if particle $i$ is in soft spot $\alpha$ and $s_{\alpha,i}=0$, otherwise.  Thus, ${\bf S}(\gamma)=\sum_\alpha {\bf s}_\alpha(\gamma)$.  Note that to calculate the soft spots we used only structural information (the particle positions and interactions).  As a result, the soft spots are structural, not dynamical features. 

We now calculate the correlation of each soft spot, ${\bf s}_\alpha(\gamma)$ with the next rearrangement at strain $\gamma_c$, ${\bf R}(\gamma_c)=\{R_i(\gamma_c)\}$, where $R_i=1$ if particle $i$ has one of the $n_\alpha$ largest displacement vectors in the critical mode and $R_i=0$ otherwise.  Here, $n_\alpha$ is the number of particles in soft spot $\alpha$.   The correlation is~\cite{Kendall}
\begin{equation}
C_\alpha^{sr}(\gamma_c-\gamma)= \frac{{\bf s}_{\alpha}(\gamma) \cdot {\bf R}(\gamma_c)}{2 n_\alpha} 
+ \frac{(1- {\bf s}_{\alpha}(\gamma)) \cdot (1 - {\bf R}(\gamma_c))}{2(N-n_{\alpha})} , \label{Csrdef}
\end{equation}
The correlation $C_\alpha^{sr}$ is unity if  ${\bf s}_{\alpha}$ and ${\bf R}$ are perfectly correlated, and zero if they are uncorrelated.  The rearrangement shown by the red arrows in Fig.~\ref{sspots} has $C_1^{sr}= 0.64$ with the blue soft spot (the ``best" soft spot with the highest value of $C^{sr}$, which we label as $\alpha=1$).

The correlation $C_1^{sr}$ depends on the number of modes, $N_m$, and number of particles per mode, $N_p$, used to define the soft spots.   We choose $N_p^{*}$ and $N_m^{*}$ to maximize the correlation $C_1^{sr}$ (Eq.~\ref{Csrdef}) with the best soft spot, averaged over all strains studied.  We find  $N_p^*=20$ particles per mode with $N_m^*=30$ modes, corresponding to roughly 13 soft spots in a 2500-particle system; however,  the results are not too sensitive to $N_p$ and $N_m$ near their maximum values as long as the fraction of particles in soft spots is fixed at $\phi_{ss} = 0.1$.

\begin{figure}
\centering \includegraphics[width=0.40\textwidth]{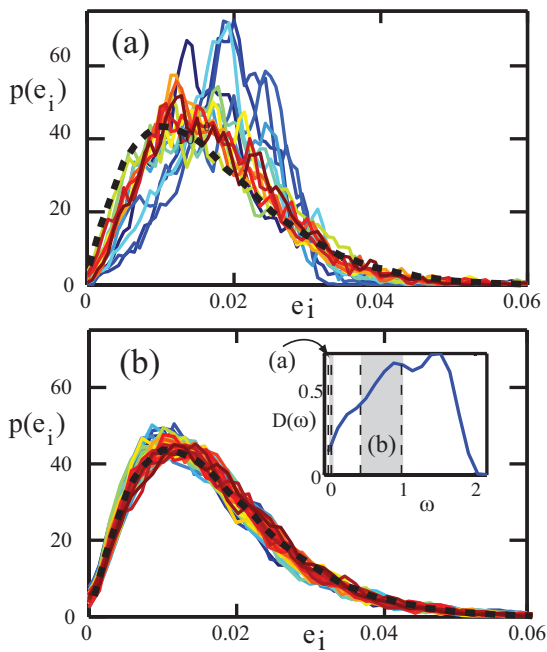}
\caption{\label{evdist} (color online) Polarization vector magnitude distributions for normal modes of a 2500 particle system.  (a) The 15 lowest frequency modes, and (b) 50 ``extended anomalous'' modes from the middle of the spectrum. The dashed line represents the distribution for a random matrix ensemble. Inset: Density of states $D(\omega)$ as a function of frequency showing the frequency ranges of the modes in (a) and (b).}
\end{figure}

What is the physical significance of the maximal values $N_m^*$ and $N_p^*$?   To understand why the $N_m^{*}=30$ lowest frequency modes are singled out, we examine the distribution of polarization vector magnitudes for each mode.  Each normal mode is composed of $N$ $d-$dimensional polarization vectors that specify the displacement of each particle in the packing.  Fig.~\ref{evdist} shows polarization vector distributions for (a) the 15 lowest frequency modes and (b) 50 intermediate frequency modes.  The location of these regimes are indicated on a plot of the density of states, $D(\omega)$ in the inset to Fig.~\ref{evdist}(a). 

Fig.~\ref{evdist}(b) shows that for modes in the middle of the spectrum, corresponding to extended anomalous modes of the type described by Wyart, et al.~\cite{Wyart2} that constitute the boson peak~\cite{Silbert}, the distributions appear to be universal with a form given by a modified Gaussian Orthogonal random matrix ensemble (solid line in Fig.~\ref{evdist}(b))~\cite{lisa}.   While most of the modes in the spectrum are well-described by this universal curve, there are clear deviations at the low and high frequency ends of the spectrum.  At the high frequency end, the localized modes differ from the universal curve but play no role in our analysis.  At the opposite end of the spectrum, Fig.~\ref{evdist}(a) shows that the lowest-frequency modes (${\cal O}(10)$ for a $2500$-particle packing) also differ significantly from the universal curve~\cite{Ning}.  This number is consistent with $N_m^{\star} \simeq 30$ and a different analysis by Schober and Oligschleger~\cite{Schober}.  

The quantity $N_p^{*}$ is the number of particles per mode in soft spots.  We estimate the size of an individual spot independently by analyzing the number of particles that change neighbors during ``elementary'' particle rearrangements. In our quasi-static simulations, ``elementary'' particle rearrangements are defined as those where the critical mode is at least 80\% correlated with the total displacement of all the particles after the packing has reached a new mechanically stable state.  The results are not sensitive to the particular threshold used as long as we exclude avalanches, in which one rearrangement triggers another, and so on. We find that the average number of particles that change neighbors during an elementary rearrangement is $10$.  This is of the same order of magnitude as $N_p \simeq 30$.  Thus, $N_m^*$ and $N_p^*$  correspond to the number of low-frequency modes that differ significantly from the extended anomalous modes, and the size of a localized rearrangement, respectively.

\begin{figure}
\centering \includegraphics[width=0.40\textwidth]{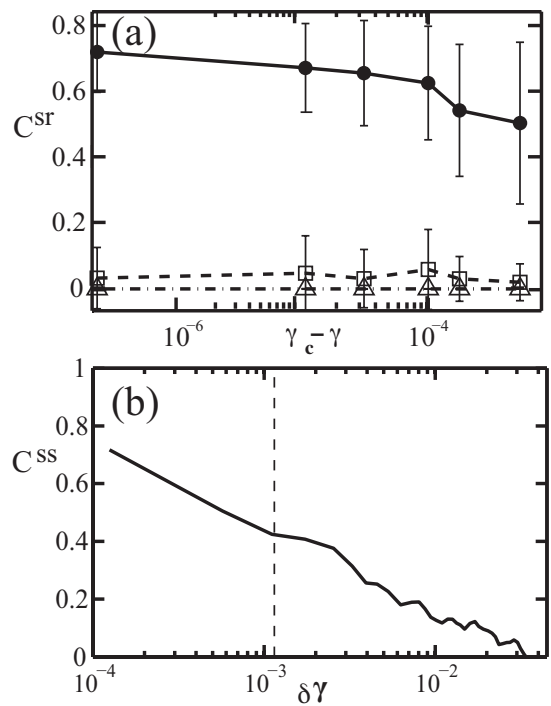}
\caption{\label{straincorr} (a) Correlation of {\em individual} soft spots with the rearrangement field $C^{sr}$ as a function of how much additional strain is required to initiate a particle rearrangement ($\gamma_c - \gamma$) for the ``best" spot with greatest overlap (solid circles), and the second (open squares) and third-ranked (open triangles) spots.  (b) Correlation of soft spot distributions as a function of the difference in strain between the distributions, $\delta \gamma$ (Eq.~\ref{Cssdef}). The vertical dashed line indicates the average strain between particle rearrangements, showing that soft spot distributions are correlated across multiple rearrangements.}
\end{figure}

Now that we have identified soft spots, we need to show that they are good candidates for structural defects, analogous to dislocations in crystalline solids, that control flow. The following properties of dislocations ensure that they control flow: (1) rearrangements tend to occur at dislocations, (2) dislocations are long-lived compared to the time between rearrangements, and (3) dislocations are structurally distinct from the rest of the system.  We now show that soft spots also possess these qualities. 
 
(1) Rearrangements occur at soft spots.  Each rearrangement is much more strongly correlated with one soft spot (the ``best'' one)  than any of the others (Fig.~\ref{straincorr}(a)).  Thus, each rearrangement occurs at one and only one soft spot in the population.  Moreover, the correlation with the best soft spot is high even when the spots are identified far in advance of the rearrangement.  In Fig.~\ref{straincorr}(a) the solid symbols show the correlation between the rearrangement and the best soft spot as a function of the difference, $\gamma_c-\gamma$, between the strain at which the rearrangement occurs, $\gamma_c$, and the strain at which the soft spot was identified, $\gamma<\gamma_c$.  The correlation decays slowly with increasing $\gamma_c-\gamma$; the best soft spot calculated shortly after a rearrangement still has a strong correlation with the next rearrangement. 

(2) The population of soft spots is long-lived compared to the interval between rearrangements.   To calculate the correlation $C_{ss}$ between soft spot distributions, we first normalize the soft spot distribution so that it has zero mean and unit variance: $\overline{\bf S}=\{ \overline S_i \}$ for all particles $i$, where $\overline{S_i} = (S_i - \sum_i S_i) / \sqrt{\sum_i S_i^2 - (\sum_i S_i)^2}$.  We then define
\begin{equation}
C_{ss}(\delta \gamma) = \frac{1}{\gamma_{tot}}\int_0^{\gamma_{tot}} d \gamma \;  {\bf \overline{S}}(\gamma) \cdot {\bf \overline{S}}(\gamma+\delta \gamma) \label{Cssdef} , 
\end{equation}
where $\gamma_{tot}$ is the total strain studied.   Fig.~\ref{straincorr}(b) shows that $C_{ss}$ decays slowly compared to the average strain between rearrangements, $\bar \gamma$.  The decay strain is approximately the product of $\bar \gamma$ and the total number of soft spots (of order 10), consistent with our observations that each rearrangement destroys a soft spot.  This would imply that $C_{ss}$ decays at a finite rate even in the thermodynamic limit where $\bar \gamma \rightarrow 0$.  

(3) The population of soft spots is structurally different from the remainder of the packing.  Compared to the rest of the system, soft spots have an average coordination number that is $6 \pm 1\%$ lower, a bond orientational order that is $32 \pm 4\%$ lower, and an excess free volume per particle that is $103 \pm  19\%$ higher.  However, we could not identify the same population of soft spots by coarse-graining these geometric quantities over the area of an average spot.  Thus, although soft spots are structurally different from the rest of the system, the difference is sufficiently subtle that one cannot identify them correctly using only these local geometric quantities.

We also calculated the local shear modulus~\cite{Tanguy1}, averaged over the area of a soft spot.  This structural quantity depends on both the local geometry and the interactions.  As first noted in~\cite{Tanguy1}, a sudden drop in the local shear modulus predicts the timing and location of the subsequent particle rearrangement.  Thus, if one is sufficiently close to the rearrangement, the spatial distribution of the coarse-grained shear modulus pinpoints when and where the rearrangement will occur.  However, it does not provide information about other soft spots that are susceptible but do not rearrange.

The soft spot analysis, on the other hand, provides fundamentally different information about the system.  Unlike the local shear modulus, the soft spot analysis identifies a {\it collection} of spots; the next rearrangement will occur at one of these spots but the analysis does not single out that particular spot {\it a priori}.

So why is it useful to identify a population of soft spots?  The advantage becomes apparent when one considers the effects of fluctuations that arise from temperature or shear.   We expect that when fluctuations are present, a rearrangement will not necessarily occur in the spot with the lowest energy barrier, but could occur in any one of the spots with some probability.   In that case, a statistical description of the soft spot population and rearrangements is necessary.

Our results show that the soft spot population, unlike the low-frequency vibrational modes from which it is derived or regions of low local shear modulus, is long-lived compared to the interval between rearrangements.  This implies that the population is robust to the fluctuations that arise in quasistatically-sheared systems.  Recent experiments on thermal colloids show that soft spots are also robust at nonzero temperatures~\cite{Chen2}.  Taken together, these results provide strong evidence that the spots do indeed constitute the structural defects relevant for flow in disordered solids. 

We thank S. R. Nagel and Ke Chen for instructive discussions.  This work was supported by DOE DE-FG02-05ER46199 (AJL).


\begin{thebibliography}{23}
\bibitem[{rem()}]{remark}
\bibinfo{howpublished}{Clusters of less than five particles are disregarded to
  avoid artifacts due to small number fluctuations.}
\end{thebibliography}
\end{document}